\documentclass[]{spie}  
\usepackage[]{graphicx}

\title{Temperature dependence of charge transfer inefficiency \\ in Chandra X-ray CCDs} 

\author{C. E. Grant, M. W. Bautz, S. E. Kissel, B. LaMarr, G. Y. Prigozhin 
\skiplinehalf
Kavli Institute for Astrophysics and Space Research, \\
Massachusetts Institute of Technology \\
Cambridge, Massachusetts 02139
}

\authorinfo{Further author information: (Send correspondence to C.E.G.)\\C.E.G.: E-mail: cgrant@space.mit.edu\\Copyright 2006 Society of Photo-Optical Instrumentation Engineers.\\This paper will be published in Proc. SPIE, vol. 6276, and is made available as an electronic preprint with permission of SPIE.  One print or electronic copy may be made for personal use only.  Systematic or multiple reproduction of any material in this paper for a fee or for commercial purposes, or modification of the content of the paper are prohibited.}

\begin{document} 
  \maketitle 

\begin{abstract}
Soon after launch, the Advanced CCD Imaging Spectrometer (ACIS), one of the focal plane instruments on the Chandra X-ray Observatory, suffered radiation damage from exposure to soft protons during passages through the Earth's radiation belts. The primary effect of the damage was to increase the charge transfer inefficiency (CTI) of the eight front illuminated CCDs by more than two orders of magnitude. The ACIS instrument team is continuing to study the properties of the damage with an emphasis on developing techniques to mitigate CTI and spectral resolution degradation. We present the initial temperature dependence of ACIS CTI from --120 to --60 degrees Celsius and the current temperature dependence after more than six years of continuing slow radiation damage. We use the change of shape of the temperature dependence to speculate on the nature of the damaging particles.
\end{abstract}


\section{INTRODUCTION}
\label{sect:intro}

The Chandra X-ray Observatory, the third of NASA's great observatories in space, was launched just past midnight on July 23, 1999, aboard the space shuttle {\it Columbia}\cite{cha2}.  After a series of orbital maneuvers Chandra reached its final, highly elliptical, orbit.  Chandra's orbit, with a perigee of 10,000~km, an apogee of 140,000~km and an initial inclination of 28.5$^\circ$, transits a wide range of particle environments, from the radiation belts at closest approach through the magnetosphere and magnetopause and past the bow shock into the solar wind.

The Advanced CCD Imaging Spectrometer (ACIS), one of two focal plane science instruments on Chandra, utilizes frame-transfer charge-coupled devices (CCDs) of two types, front- and back-illuminated (FI and BI).  Soon after launch it was discovered that the FI CCDs had suffered radiation damage from exposure to soft protons scattered off the Observatory's grazing-incidence optics during passages through the Earth's radiation belts\cite{gyp00,odell}.  Since mid-September 1999, ACIS has been protected during radiation belt passages and there is an ongoing effort to prevent further damage and to develop hardware and software strategies to mitigate the effects of charge transfer inefficiency on data analysis\cite{cticorr}.

One symptom of radiation damage in CCDs is an increase in the number of charge traps.  When charge is transfered across the CCD to the readout, some portion can be captured by the traps and gradually re-emitted.  If the original charge packet has been transfered away before the traps re-emit, the captured charge is ``lost'' to the charge packet. The pulseheight read out from the instrument which corresponds to a given energy decreases with increasing transfer distance.  This process is quantified as charge transfer inefficiency (CTI), the fractional charge loss per pixel and is calculated from a linear fit to the pulseheight versus row number; CTI = (slope/intercept).  

Damage can exist in the imaging or framestore array, causing parallel CTI in the column direction, or in the serial register, causing serial CTI along rows.  The ACIS CCDs come in two flavors, front- and back-illuminated, which have different manifestations of CTI.  The distribution of re-emission time constants of the electron traps which cause CTI vary depending on the type of damage.  In addition, the pixel to pixel transfer time in the imaging, framestore and serial arrays differs so that the same species of electron trap can produce different CTI results in different locations.

The eight front-illuminated CCDs had essentially no CTI before launch, but are strongly sensitive to radiation damage from low energy protons ($\sim$100~keV) which preferentially create traps in the buried transfer channel.  The framestore covers are thick enough to stop this radiation, so the initial damage was limited to the imaging area of the FI CCDs.  Radiation damage from low-energy protons is now minimized by moving the ACIS detector away from the aimpoint of the observatory during passages through the Earth's particle belts.  Continuing exposure to both low and high energy particles over the lifetime of the mission slowly degrades the CTI further.\cite{odell,ctitrend}  As of January 2000, the parallel CTI at 5.9~keV of the ACIS FI CCDs varied across the focal plane from $1 - 2 \times 10^{-4}$ at the nominal operating temperature of --120$^\circ$~C with a rate of increase of roughly $3 \times 10^{-6}$/year.  Parallel CTI in the framestore array and serial CTI were not affected by the initial radiation damage and remain negligible with upper limits of $< 10^{-6}$ (framestore array) and $< 2 \times 10^{-5}$ (serial array).

The two back-illuminated CCDs (ACIS-S1,S3) suffered damage during the manufacturing process and exhibit CTI in both the imaging and framestore areas and the serial transfer array, but are less sensitive to the low energy particles which damage the FI CCDs because they cannot reach the transfer channel.  The parallel CTI at 5.9~keV of the S3 BI CCD was  $\sim 1.6 \times 10^{-5}$ at a temperature of --120$^\circ$~C at the beginning of the mission with a strong non-linear flattening of pulseheight at low row numbers due to CTI in the framestore array.  The serial CTI is much larger, $\sim 8 \times 10^{-5}$. BI CCD parallel CTI has been increasing at a rate of $1 \times 10^{-6}$.

Measured CTI is a function of temperature.  Detrapping time constants decrease as the temperature increases so that different populations of traps can become more or less important.  If the detrapping time constant drops below the pixel transfer time or becomes much longer than the typical distance between charge packets, charge is no longer lost to the trap.  The distribution of trap time constants at a particular temperature determines the CTI, so temperature can positively or negatively correlate with CTI.  Conversely, the temperature dependence of CTI reflects the particular blend of electron traps.  The temperature dependence of ACIS CTI was first measured in 1999 \cite{gyp00} and again in 2005.  In this paper, we present a comparison of the CTI temperature dependence in both time periods and use this to speculate as to the nature of the intervening particle damage.

This paper begins by describing the calibration data taken during the temperature tests in Section~\ref{sect:data} and the CTI data reduction in Section~\ref{sect:reduc}.  The CTI-temperature dependence for both time periods is presented and discussed in Section~\ref{sect:results}.

\section{DATA}
\label{sect:data}

\begin{figure}
\vspace{3.9in}
\includegraphics{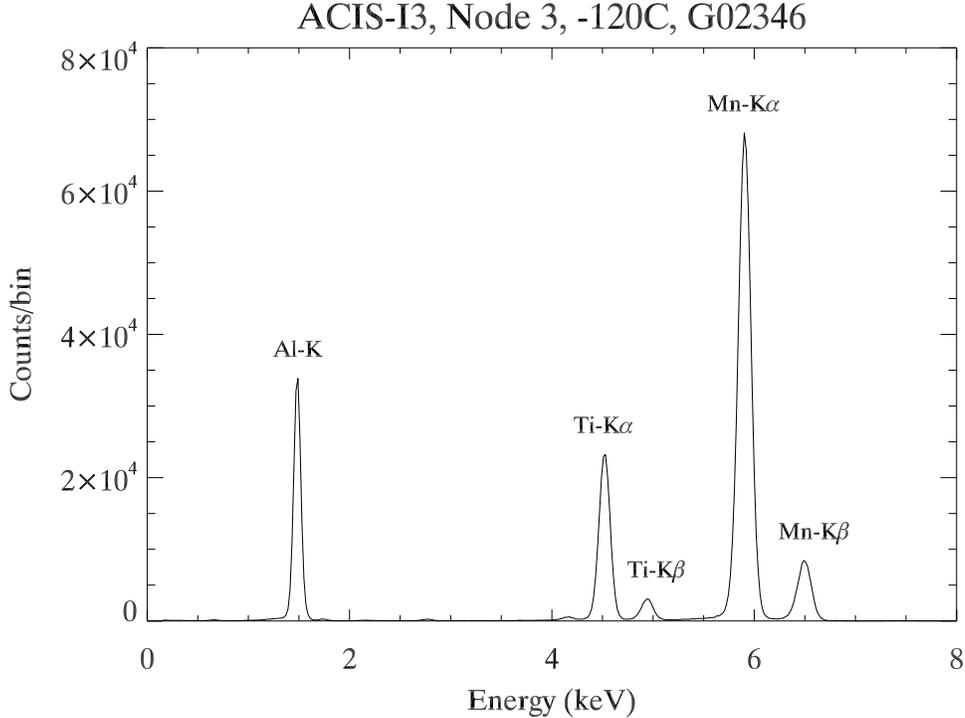}
\caption{The spectrum of the ACIS calibration source on an FI CCD.  Only data from the first 256 rows are included to minimize degradation by CTI.  The strongest spectral features are labelled.}
\label{fig:ecsspectrum}
\end{figure}

\begin{figure}
\vspace{3.9in}
\includegraphics{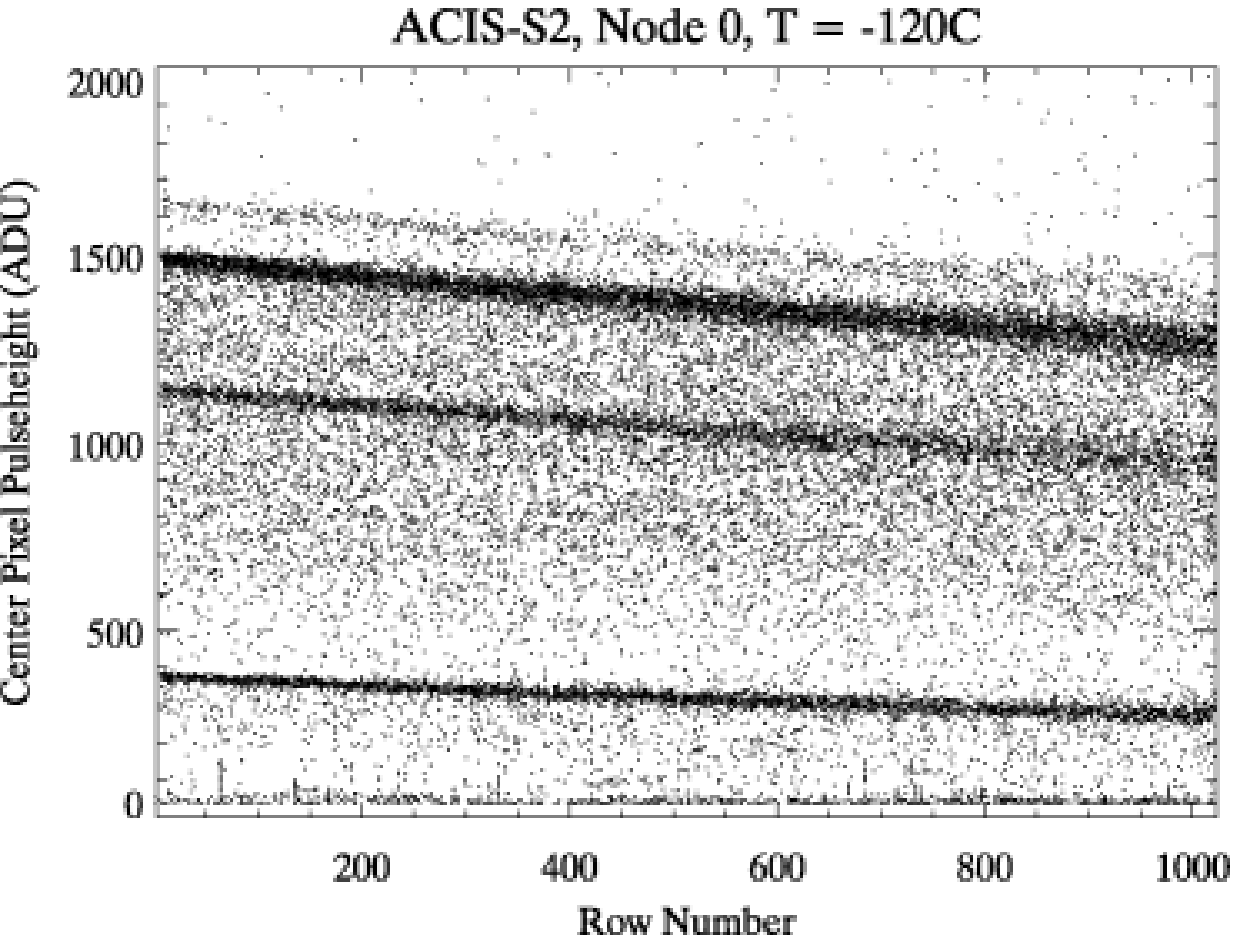}
\caption{A scatter plot of the center pixel pulseheight of each X-ray event versus its row number for a single node of the FI CCD S2 at the nominal operating temperature of --120$^\circ$~C.  The diagonal lines are spectral features in the ACIS calibration source.  Events split over multiple pixels appear at lower pulseheights than equivalent single pixel events.}
\label{fig:s2cti}
\end{figure}

All the results presented here are based on observations of the ACIS External Calibration Source (ECS) which uniformly illuminates the ACIS CCDs when ACIS is moved out of the telescope focus.  The ECS consists of $^{55}$Fe with Al and Ti fluorescence targets.  ACIS is placed in the stowed position exposing the CCDs to the ECS which produces many spectral features, the strongest of which are Mn-K$\alpha$ (5.9~keV), Ti-K$\alpha$ (4.5~keV), and Al-K (1.5~keV).  Figure~\ref{fig:ecsspectrum} shows a spectrum of the ECS and labels the brightest features.   Figure~\ref{fig:s2cti} is a scatter plot of the center pixel pulseheight of each X-ray event versus its row number for a single node of the FI CCD S2 at the nominal operating temperature of --120$^\circ$~C.  The data are taken in the standard Timed Exposure mode with a 3.2~second frame time.  X-ray events on ACIS CCDs can produce charge packets which occupy multiple pixels and so are recorded as 3 x 3 pixel event islands.  We compute CTI from the center pixel pulseheight alone.

The 1999 dataset consists of 8 observations taken during a single orbit on 17-19 September 1999.  The focal plane temperature ranges from --109.2$^\circ$~C to --69.7$^\circ$~C.  An additional data point from 7 October 1999 with a focal plane temperature of --118.9$^\circ$~C is included to increase the temperature range.  There are no unprotected radiation belt passages between this later point and the earlier data and the time differences is short enough that we do not expect any significant change in accumulated radiation damage. The exposure time ranges from 3.2 to 12.0~ksecs with a mean of 5.1~ksec. Only events from the front illuminated (FI) CCD S2 and and the back-illuminated (BI) CCD S3 were telemetered.  Otherwise, the operating mode is nearly identical to that currently used for CTI monitoring. 

The 2005 dataset consists of 45 standard CTI monitoring observations taken from 24 July through 5 October 2005. The focal plane temperature ranges from --121.0$^\circ$~C to --89.7$^\circ$~C. These observations utilize all ten of the ACIS CCDs, but only data from S2 and S3 are analyzed for comparison with the 1999 data.

Measured CTI is a function of fluence or, more specifically, the amount of charge deposited on the CCD.\cite{gendreau}  As the fluence increases, traps filled by one charge packet may remain filled as a second charge packet is transferred through the pixel.  The second charge packet sees fewer unoccupied traps as a result of the previous ``sacrificial charge'' and loses less charge then it would have otherwise.  Whether sacrificial charge is important depends on the interaction of the pixel transfer time, the typical distance between charge packets, and the re-emission time constant of the traps.  For observations of the ACIS ECS and for most Chandra observations in general, the primary source of sacrificial charge is the particle background.  The measured CTI will therefore be a function of the particle environment. (For more on sacrificial charge in ACIS CCDs, see Ref.~\citenum{saccharge})

Fortunately, the particle background levels during the 1999 and 2005 datasets are relatively similar so variations due to sacrificial charge should be minimal.  As a measure of the particle background rate in situ, we are using the counting rate of events on the BI CCD S3 with pulseheights greater than 3750~ADU ($\sim 15$~keV) which correlates well with high energy ($>$ 10 MeV) cosmic ray protons \cite{bkg}.  At these high energies, the effective area of the telescope mirrors is essentially zero, so these events are caused by energetic particles rather than astrophysical X-rays.  The mean particle background levels in 1999 and 2005 are similar, but not identical and the variance in the 2005 data is much higher due to larger scatter and a solar event.  We have further selected the 2005 observations to better match the 1999 particle background.  Of the original 45 observations, 31 have acceptable background rates.  The mean particle background levels as measured by the S3 high energy reject rate was 74 and 72 cts/frame in 1999 and 2005 respectively, with a standard deviation of 1.0 and 3.1 cts/frame.  No correction is being made for sacrificial charge, however we add an additional error term in quadrature to account for the variance in the background rate. 

After filtering, the exposure times for the 2005 observations range from 5.5 to 8.2~ksec with a mean of 7.7~ksec.  While these exposure times are comparable to the 1999 observations (3.2 to 12.0~ksec, mean of 5.1~ksec), the ACIS ECS consists of radioactive $^{55}$Fe which decays with a half-life of 2.73~years.  Therefore, the source in 1999 was 4.5 times brighter than in 2005.  Given the similar background level, this implies that the signal to noise is similarly much worse in 2005 than in 1999.

\section{DATA REDUCTION}
\label{sect:reduc}

Both the 1999 and 2005 datasets were analyzed identically. In general, the 1999 dataset has slightly larger eventlists and much higher signal to noise due to the radioactive decay of the ECS discussed above.  Measuring parallel CTI is relatively straight-forward; simply fit a linear function to the pulseheight as a function of row.  Serial CTI is measured in much the same way, using column instead of row number, but can be difficult to disentangle from parallel CTI effects.  We measure serial CTI by selecting events from the bottom rows of the CCD, thus minimizing any crosstalk with parallel CTI.  The number of events used to measure serial CTI is therefore much lower than for parallel CTI and the statistical errors are much larger.  All events are included in FI CCD CTI fitting while the S3 BI CCD datasets were filtered to include only single pixel and vertically split events.  This lowers the background level and allows for more reliable fits of the spectral line.  A small fraction of source events are lost to the filtering, but these events appear to be uniformly distributed across the CCD and should not bias our CTI measurements.

\begin{figure}
\vspace{3.9in}
\includegraphics{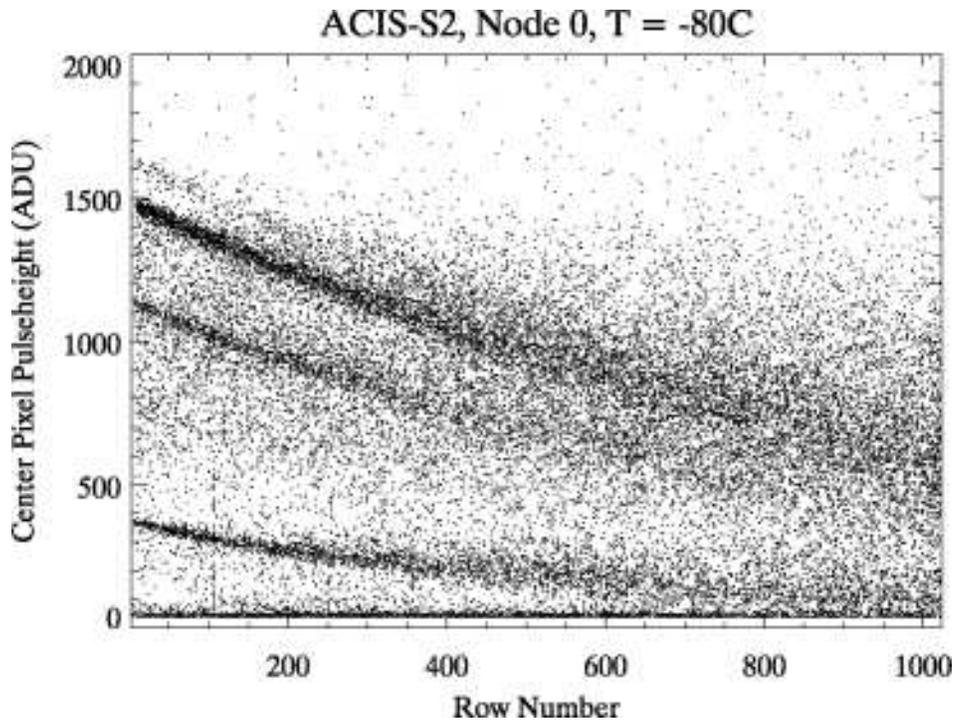}
\caption{A scatter plot of the center pixel pulseheight of each X-ray event versus its row number for a single node of the FI CCD S2 at a temperature of --80$^\circ$~C. Comparison to Figure~\ref{fig:s2cti} shows the additional CTI degradation at high temperatures for the FI CCDs.}
\label{fig:s2cti70}
\end{figure}

For the S2 FI CCD at temperatures above --110$^\circ$~C, the Mn-K$\alpha$ line (5.9~keV) is increasingly blended with with the neighboring Ti-K$\alpha$ line (4.5~keV) at high row numbers where the effects of CTI are worst.  Figure~\ref{fig:s2cti70} is a scatter plot of center pixel pulseheight for the S2 FI CCD at --80$^\circ$~C and can be compared to the --120$^\circ$~C data shown in Figure~\ref{fig:s2cti}.  Our standard automated CTI processing algorithm is unable to reliably fit the broadened and blended line. Instead, the pulseheight versus row number was fit in Event Browser\footnote{Event Browser is part of a package of data analysis tools developed by the ACIS team at Penn State.  TARA: Tools for ACIS Review and Analysis is available from http://www.astro.psu.edu/xray/docs/TARA} using a linear regression with sigma clipping algorithm with the initial values input by hand. This seems to produce more reasonable and more repeatable results for the high temperature data. For consistency, all the S2 data was processed in this manner. At low temperatures, the absolute CTI is not identical to the standard data processing, but has a small offset of about $6 \times 10^{-6}$.  The linear regression algorithm does not output a statistical error on the fitted parameters, so we repeated each fit five times with different initial conditions and used the standard deviation to characterize the parameter error.

\section{RESULTS}
\label{sect:results}

\begin{figure}
\vspace{3.9in}
\includegraphics{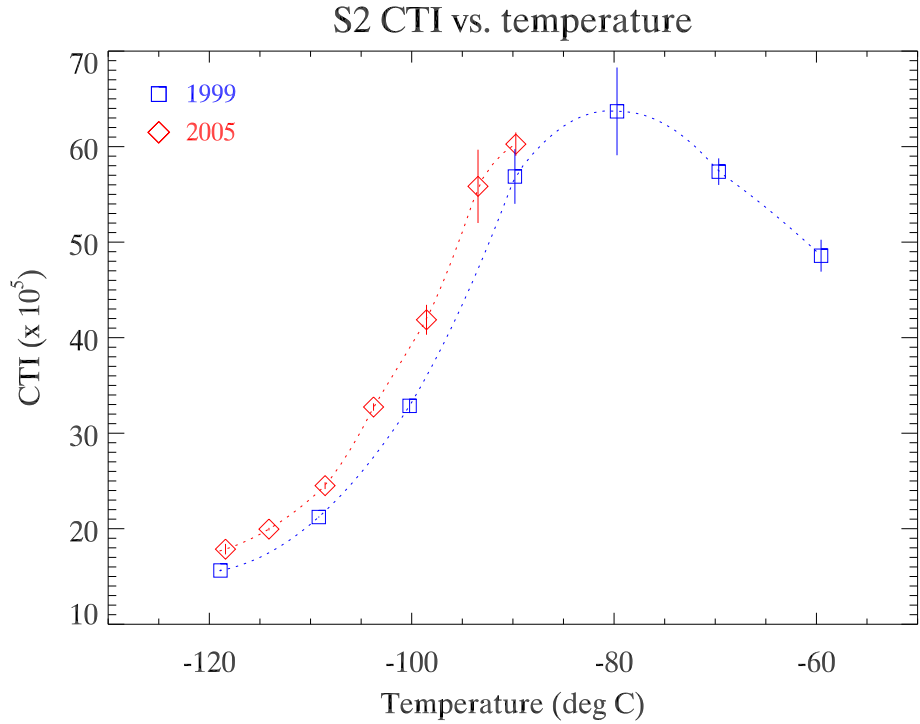}
\caption{S2 FI CCD 5.9~keV parallel CTI versus temperature.  The squares and diamonds indicate data taken in 1999 and 2005, respectively.  The dotted line is an interpolation intended to highlight the curvature. }
\label{fig:s2ctitemp}
\end{figure}

\begin{figure}
\vspace{3.9in}
\includegraphics{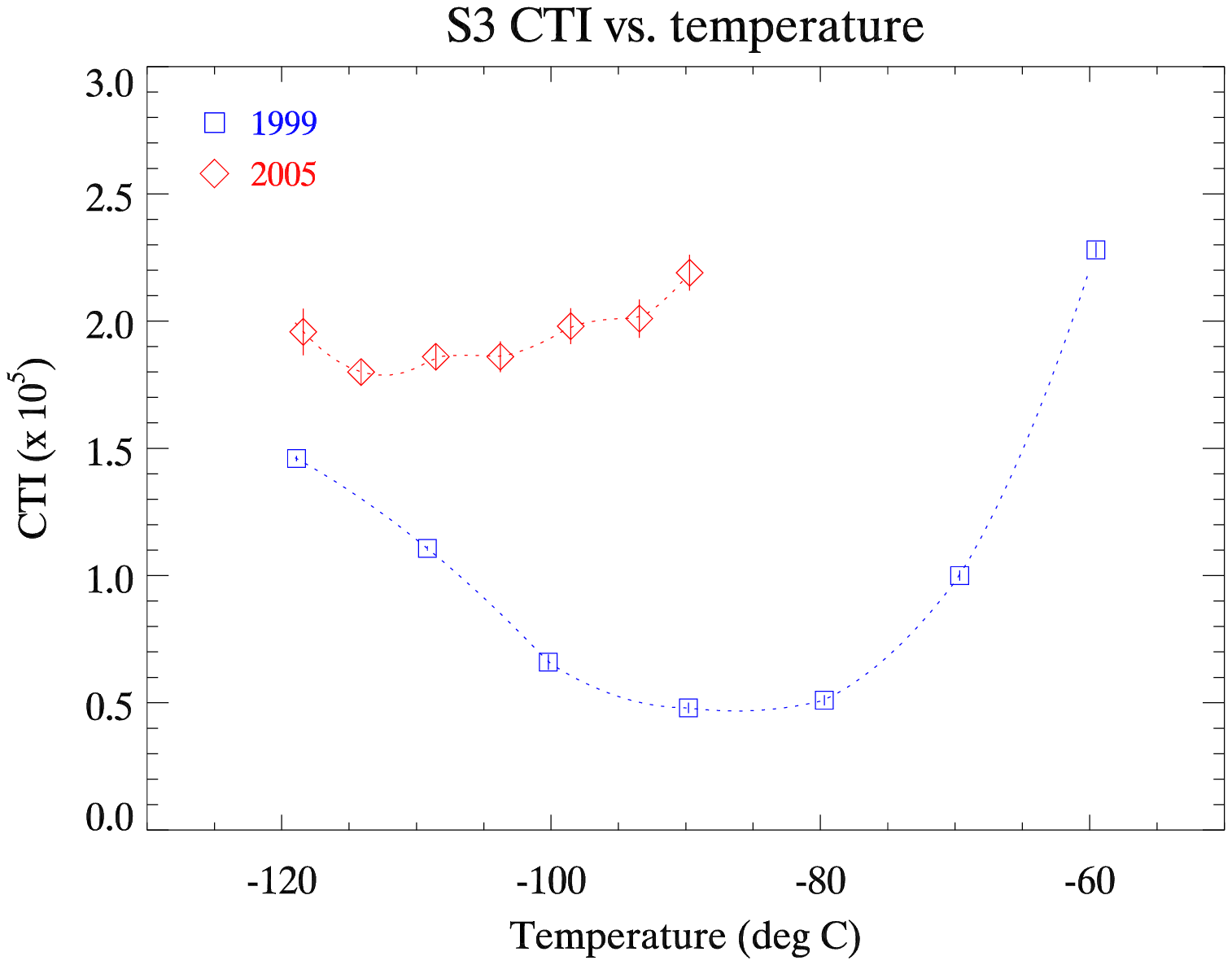}
\caption{S3 BI CCD 5.9~keV parallel CTI versus temperature.  The squares and diamonds indicate data taken in 1999 and 2005, respectively.  The dotted line is an interpolation intended to highlight the curvature.}
\label{fig:s3ctitemp}
\end{figure}

\begin{figure}
\vspace{3.9in}
\includegraphics{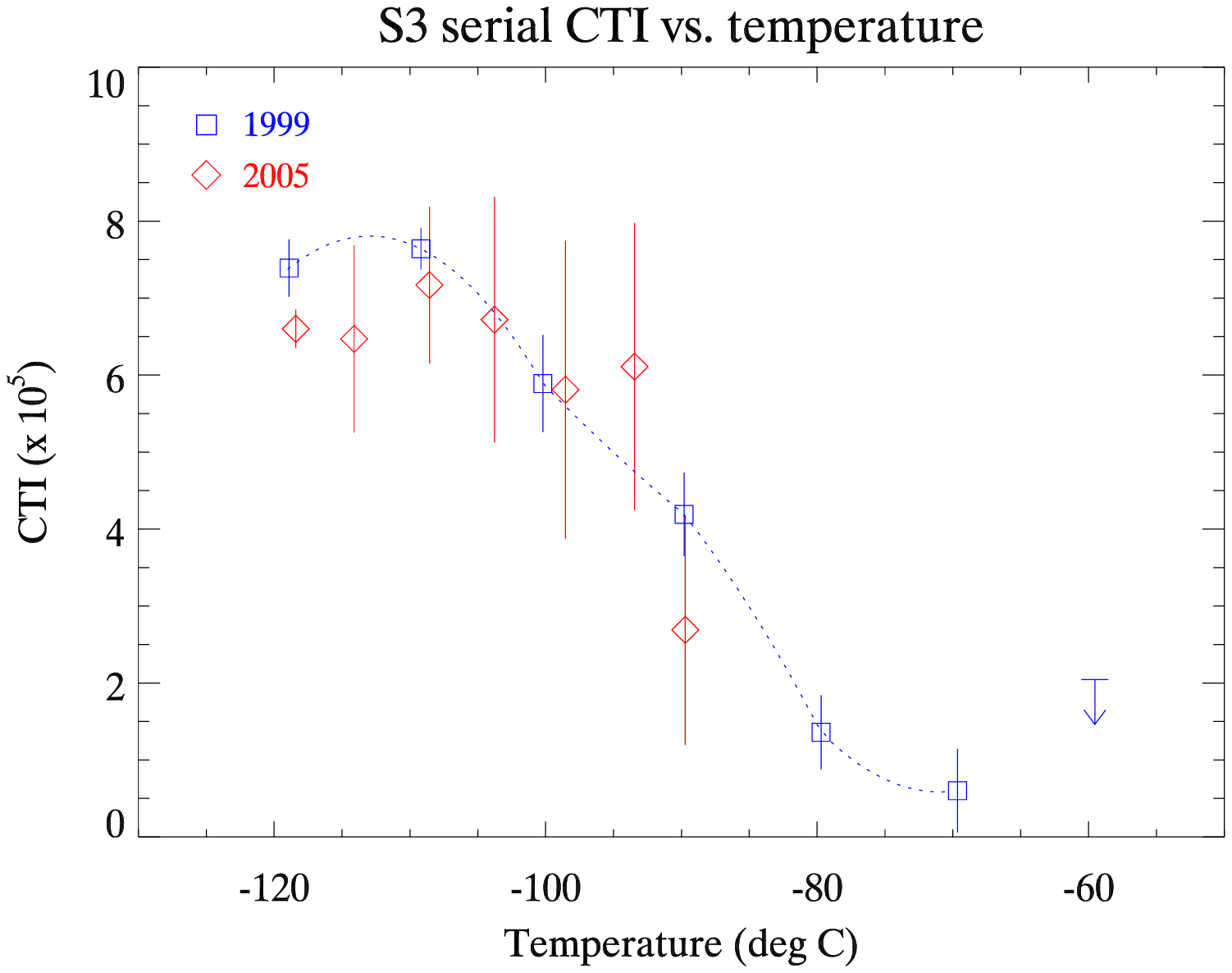}
\caption{S3 BI CCD 5.9~keV serial CTI versus temperature.  The squares and diamonds indicate data taken in 1999 and 2005, respectively.  The dotted line is an interpolation intended to highlight the curvature.  The upper limit is 3-sigma.}
\label{fig:s3ctistemp}
\end{figure}

The measured CTI as a function of temperature is shown in Figures~\ref{fig:s2ctitemp} through \ref{fig:s3ctistemp}.  The parallel CTI of the FI CCD S2 is in Figure~\ref{fig:s2ctitemp}, while Figures~\ref{fig:s3ctitemp} and \ref{fig:s3ctistemp} are the parallel and serial CTI of the BI CCD S3.  The serial CTI of the FI CCDs remains below our measurement errors.  The squares and diamonds indicate data taken in 1999 and 2005, respectively and the error bars (1-sigma, 68\% confidence), include both the statistical measurement error and an estimate of the additional systematic error.  The 2005 low temperature data has been binned for clarity.  The dotted line is not a fit but is a three point quadratic interpolation.  It is intended only to guide the eye and provide an estimate of the direction of curvature.

In both cases the parallel CTI has increased between 1999 and 2005 while the BI CCD serial CTI is consistent with no change albeit with large error bars. The size of the increase at low temperatures is consistent with that measured during regular CTI monitoring\cite{ctitrend}.  In both cases the shape of the parallel CTI-temperature relation has changed as well. This is not unexpected since the cause of the initial damage (FI CCD S2: low energy protons in the radiation belts, BI CCD S3: manufacturing) is not identical to the cause of the slow increase in the six intervening years (low and high energy protons from solar storms and cosmic rays). The change in the parallel CTI-temperature dependence is most impressive for the BI CCD S3.  In this case the initial CTI in 1999 was small so the increase due to continual radiation damage is a larger fraction of the total. For the FI CCD S2 the change in the CTI/temperature dependence is much more subtle with only a slight change in the curvature.

In our previous work\cite{ctitrend}, we determined that the initial damage and the ongoing accumulated damage did not have the same particle spectrum based primarily on the differences between the BI and FI CCD CTI change.  The initial damage which affected only the FI CCDs should be due to softer particles than the slower accumulated damage which affects both.  That the change in the FI CCDs is larger than the BI CCDs implies that the particle spectrum must extend to lower energies below 1~MeV.

\begin{figure}
\vspace{4.0in}
\includegraphics{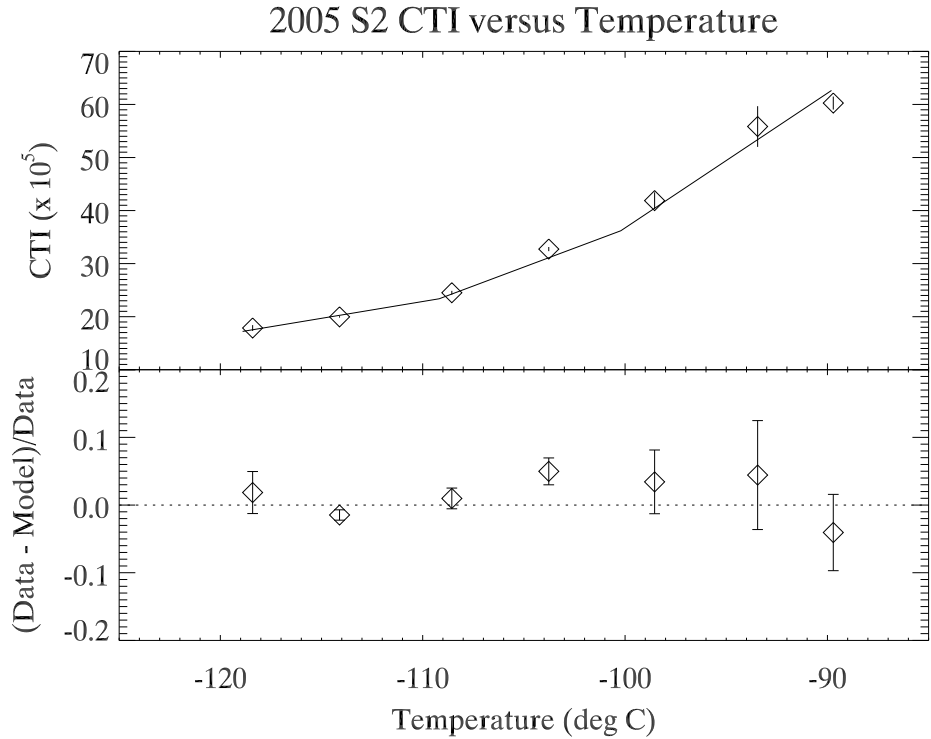}
\caption{The 2005 CTI-temperature dependence for the FI CCD S2 as shown previously in Figure~\ref{fig:s2ctitemp}.  The solid line is a model for the data consisting of the 1999 S2 data multiplied by a scaling factor.}
\label{fig:s2model}
\end{figure}

We can also use the change in the temperature dependence of CTI to infer general properties of the damaging particle spectrum.  The CTI-temperature dependence of the FI CCD S2 in 2005, after six years on orbit, is shown again in Figure~\ref{fig:s2model}.  This data is reasonably well fit by a model which is simply the 1999 S2 data multiplied by a scale factor.  The model assumes that the ongoing radiation damage produces the same CTI-temperature dependence as the initial low energy proton damage.

\begin{figure}
\vspace{4.0in}
\includegraphics{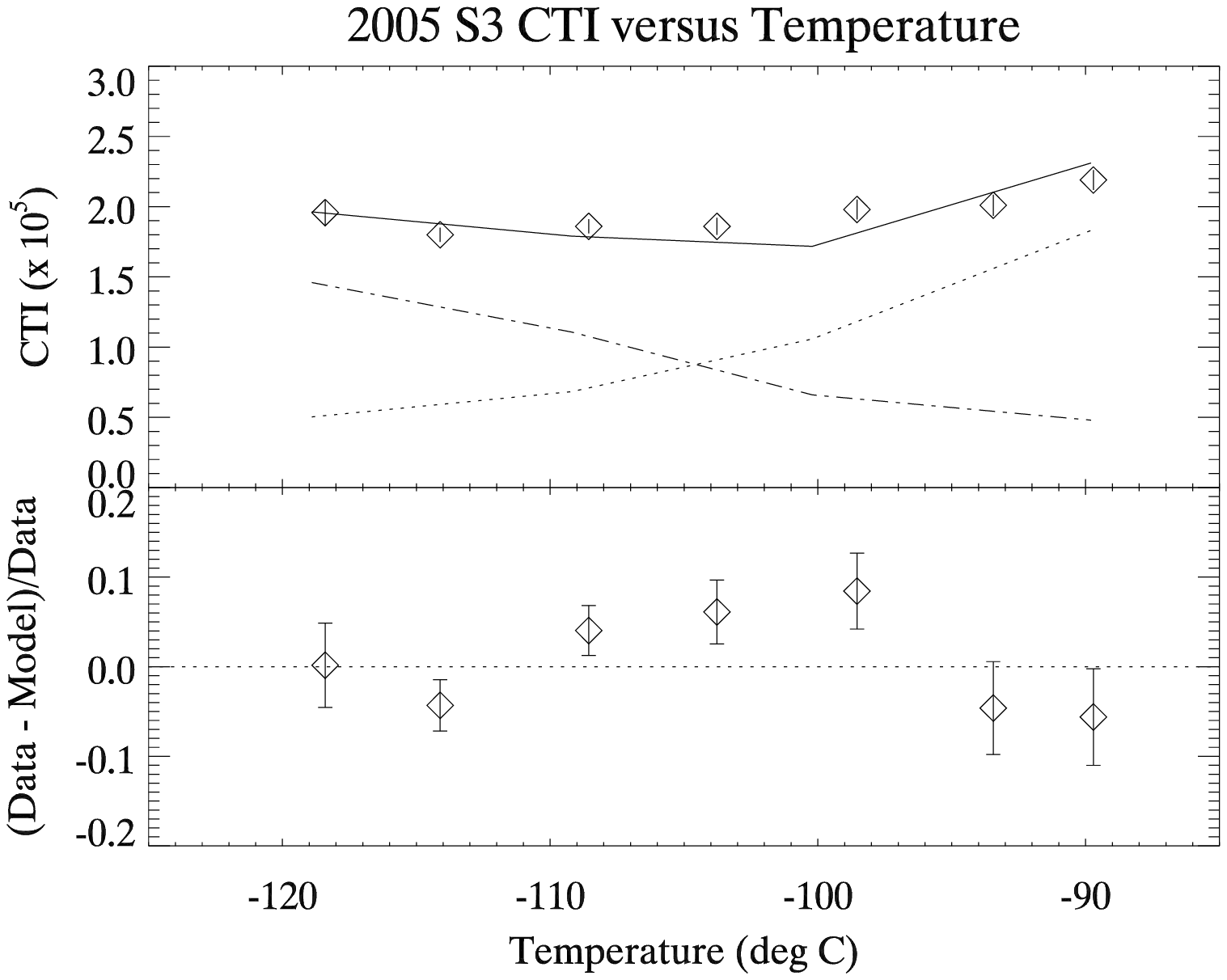}
\caption{The 2005 CTI-temperature dependence for the BI CCD S3 as shown previously in Figure~\ref{fig:s3ctitemp}.  The solid line is a model for the data consisting of the sum of the 1999 S2 data multiplied by a scaling factor (dotted) and the 1999 S3 data (dot-dash).}
\label{fig:s3model}
\end{figure}

Similarly, Figure~\ref{fig:s3model} compares the CTI-temperature dependence of the BI CCD S3 in 2005 to a model.  In this case, the model consists of the sum of the 1999 S3 data, to represent the initial manufacturing damage, and the scaled 1999 S2 data, to represent the ongoing radiation damage accumulation. The ratio of the S2 and S3 scaling factors is $\sim$3.1 which is in agreement with the ratio of the measured CTI increases.\cite{ctitrend}  Again the data is reasonably well fit by this model.  In both cases, however, the deviations from the model appear non-random and the temperature dependence of the deviations is similar.  This may be further evidence that the particle spectrum that caused the initial damage is not the same as the ongoing damage particle spectrum. 

\section{Conclusion}

The CTI-temperature dependence of the Chandra ACIS CCDs has been measured twice in orbit; soon after launch in 1999 and again in 2005.  We presented analysis of this data for back-illuminated and front-illuminated CCDs and discussed the differences between the two CCDs and between the two sets of observations.  The temperature dependence can be used to characterize the particular electron trap distribution.  In the case of ACIS, the initial temperature dependence of the FI and BI CCDs in 1999 was very different, due to the different causes of the damage.  The change in the temperature dependence between 1999 and 2005 was similar for both the FI and BI CCDs indicating that the spectrum of particle energies must be hard enough to reach the BI CCD buried channel.  Small differences between the change in temperature dependence in 1999 and 2005 and the initial FI CCD temperature dependence may indicate that the traps induced by the low energy protons in the radiation belts are slightly different than the traps created by the harder continuing particle damage.

\acknowledgments
The authors would like to acknowledge the work of Leisa Townsley and Pat Broos at Penn State University in developing and maintaining a number of highly useful data analysis tools for X-ray astronomy.  This work was supported by NASA contracts NAS 8-37716 and NAS 8-38252.


\bibliography{paper}
\bibliographystyle{spiebib}

\end{document}